\documentclass[a4paper,11pt]{article}
\usepackage{graphicx,epsfig}
\usepackage{fancyhdr,fancybox}
\usepackage{indentfirst}
\usepackage{verbatim}
\usepackage[sort&compress, numbers]{natbib}
\usepackage{geometry}
\usepackage{extarrows} % 画公式中的箭头
\usepackage{chemarrow}
\usepackage[small]{caption2}
\usepackage{color}
\usepackage{microtype}
\DisableLigatures[f]{encoding = *, family = *}

\usepackage{times}     % 包括 Times Roman + Helvetica + Courier
\usepackage{amssymb}                 % 用来排版漂亮的数学公式
\usepackage{amsthm}
\usepackage{mathrsfs}                % 英文花体字体
\usepackage{bbm}
\usepackage{hyperref}  % 生成书签

\newcommand{\paperfont}{\fontsize{12pt}{1.3\baselineskip}\selectfont}
\begin{document}
%\linenumbers

%%%%%%%%%% 定理类环境的定义 %%%%%%%%%%
\theoremstyle{definition}
\makeatletter
\thm@headfont{\bf}
\makeatother
\newtheorem{definition}{Definition}
\newtheorem{example}{Example}
\newtheorem{theorem}{Theorem}
\newtheorem{lemma}{Lemma}
\newtheorem{corollary}{Corollary}
\newtheorem{remark}{Remark}
\newtheorem{proposition}{Proposition}

%%%%%%%%%% 页眉和页脚的设置 %%%%%%%%%%
\lhead{}
\rhead{}
\lfoot{}
\rfoot{}

%%%%%%%%%% 一些重定义 %%%%%%%%%%
\renewcommand{\refname}{References}
\renewcommand{\figurename}{Figure}
\renewcommand{\tablename}{Table}
\renewcommand{\proofname}{Proof}

%%%%%%%%%% 符号重定义 %%%%%%%%%%
\newcommand{\diag}{\mathrm{diag}}
\newcommand{\one}{\mathbbm{1}}

%%%%%%%%%% 方程按章节编号 %%%%%%%%%%
%\numberwithin{equation}{section}

%%%%%%%%%% 论文标题、作者等 %%%%%%%%%%
\title{\textbf{Relaxation rates of gene expression kinetics reveal the feedback signs of autoregulatory gene networks}}
\author{Chen Jia$^1$, Hong Qian$^2$, Min Chen$^{1,*}$, Michael Q. Zhang$^{3,4,*}$ \\
\footnotesize $^1$Department of Mathematical Sciences, University of Texas at Dallas, Richardson, TX 75080, USA\\
\footnotesize $^2$Department of Applied Mathematics, University of Washington, Seattle, WA 98195, USA\\
\footnotesize $^3$Department of Biological Sciences, Center for Systems Biology, University of Texas at Dallas, Richardson, TX 75080, USA\\
\footnotesize $^4$MOE Key Lab and Division of Bioinformatics, CSSB, TNLIST, Tsinghua University, Beijing 100084, China\\
\footnotesize $^*$Correspondence: mchen@utdallas.edu (M.C.), michael.zhang@utdallas.edu (M.Q.Z.)}
\date{}                              % 日期
\maketitle                           % 生成标题
%\tableofcontents                    % 插入目录
\thispagestyle{empty}                % 首页无页眉页脚

%%%%%%%%%% 正式使用字体 %%%%%%%%%%%
\paperfont

%%%%%%%%%% 摘要 %%%%%%%%%%
\begin{abstract}
The transient response to a stimulus and subsequent recovery to a steady state are the fundamental characteristics of a living organism. Here we study the relaxation kinetics of autoregulatory gene networks based on the chemical master equation model of single-cell stochastic gene expression with nonlinear feedback regulation. We report a novel relation between the rate of relaxation, characterized by the spectral gap of the Markov model, and the feedback sign of the underlying gene circuit. When a network has no feedback, the relaxation rate is exactly the decaying rate of the protein. We further show that positive feedback always slows down the relaxation kinetics while negative feedback always speeds it up. Numerical simulations demonstrate that this relation provides a possible method to infer the feedback topology of autoregulatory gene networks by using time-series data of gene expression.

\noindent \\% 不缩进
\end{abstract}

%%%%%%%%%% 正文 %%%%%%%%%%
\section{Introduction}
In recent years, it has become increasingly clear that stochastic effects play a crucial role in molecular biology. Growing numbers of single-cell measurements, some with single-molecule resolution, have revealed that gene expression is a complex stochastic process involving various probabilistic chemical reactions \cite{mcadams1997stochastic, elowitz2002stochastic, ozbudak2002regulation,blake2003noise, raser2004control, kaern2005stochasticity, raser2005noise, golding2005real, raj2006stochastic, xie2008single}. With the effort of many scholars, significant progress has been made in the stochastic biochemical reaction kinetics of gene regulatory networks, which has a dual representation in terms of its probability distribution and stochastic trajectory. The former is usually described by a chemical master equation, first introduced in the work of Leontovich \cite{leontovich1935basic} and Delbr\"{u}ck \cite{delbruck1940statistical}, while the latter is the solution to a stochastic integral equation with time-changed Poisson representation \cite{anderson2015stochastic} that can be computationally
simulated using Gillespie's algorithm.

It is a fundamental question of how the topology, especially the feedback mechanism, of a gene regulatory network can affect the behavior of stochastic gene expression and to what extent the topological information can be extracted from the massive and noisy data of gene expression. Systems biology has been trying to understand the complexity of biochemical networks by identifying characteristic properties of common network motifs \cite{milo2002network, hilfinger2016exploiting}. In fact, the steady-state behavior of stochastic gene expression has been extensively studied over the past two decades \cite{paulsson2000stochastic, paulsson2000random, kepler2001stochasticity, karmakar2004graded, walczak2005absolute, friedman2006linking, shahrezaei2008analytical, taniguchi2010quantifying, assaf2011determining, kumar2014exact, petrosyan2014oscillations, jia2014modeling, pajaro2015shaping, lin2016gene, pajaro2017stochastic, bressloff2017stochastic, jia2017simplification, jia2017emergent}. It has been shown from both the experimental \cite{becskei2001positive, becskei2000engineering, shimoga2013synthetic} and theoretical \cite{jia2017stochastic} aspects that positive feedback amplifies noise in gene expression, while negative feedback reduces such noise. However, studies on the time-dependent dynamic behavior of stochastic gene expression have been quite limited, although it has been shown that a gene circuit with negative autoregulation has an accelerated transcriptional response \cite{rosenfeld2002negative}.

From the viewpoint of biochemical kinetics, the relaxation of a gene regulatory network in a single cell, such as the recovery of a sensory system to the steady state in response to an input stimulus, is both theoretically fundamental and practically informative. Experiments have indicated that the relaxation time varies significantly among different genes in different species \cite{belle2006quantification, eden2011proteome, moran2013sizing, christiano2014global}, from prokaryotes to yeast then to higher eukaryotes. In the present work, we study the relaxation kinetics for stochastic gene expression in autoregulatory gene networks.  We are particularly interested in the relationship between the relaxation rate and the feedback topology of the underlying gene circuit.

% Results and Discussion can be combined.
\section{Model}
\begin{figure}[!htb]
\begin{center}
\centerline{\includegraphics[width=0.6\textwidth]{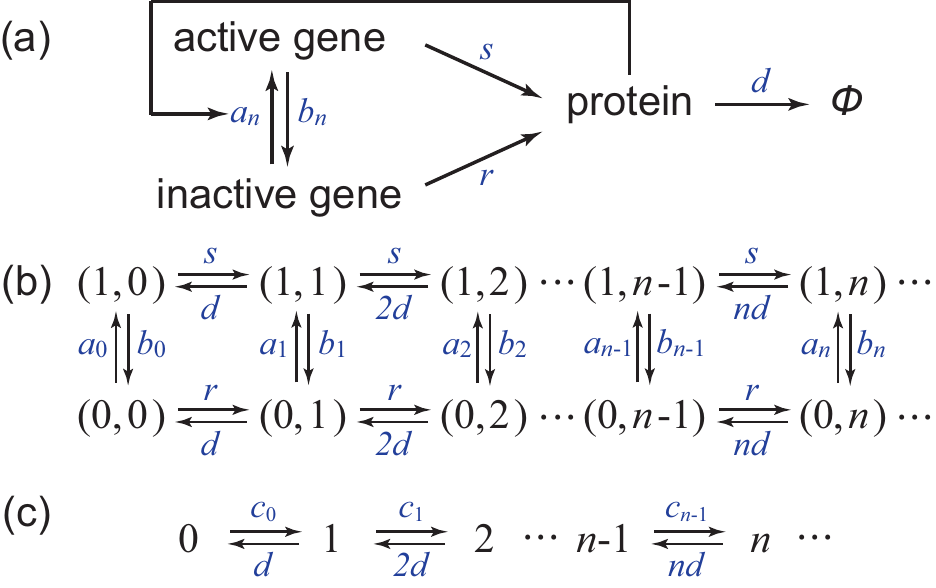}}
\caption{\textbf{Models of stochastic gene expression in autoregulatory gene networks.} (a) A minimal autoregulatory gene network with positive or negative feedback. (b) The transition diagram of the DGP. The biochemical state of the gene is described by $(i,n)$, where $i = 0,1$ represents the gene activity and $n = 0,1,2,\cdots$ represents the protein copy number. (c) The transition diagram of the reduced Markov model when the gene switches rapidly between the active and inactive states. In the reduced model, the biochemical state of the gene is only described by the protein copy number $n$.}\label{model}
\end{center}
\end{figure}

We consider an autoregulatory gene network in a single cell involving the following reactions, as also illustrated in Fig. \ref{model}(a) \cite{peccoud1995markovian, hornos2005self, shi2011perturbation, grima2012steady, potoyan2015dichotomous, ge2015stochastic, liu2016decomposition}:
\begin{gather*}
\textrm{inactive gene} \xlongrightarrow{a_n} \textrm{active gene},\\
\textrm{active gene} \xlongrightarrow{b_n} \textrm{inactive gene},\\
\textrm{active gene} \xlongrightarrow{s} \textrm{active gene}+\textrm{protein},\\
\textrm{inactive gene} \xlongrightarrow{r} \textrm{inactive gene}+\textrm{protein},\\
\textrm{protein} \xlongrightarrow{d} \varnothing,
\end{gather*}
where the first two reactions characterize the switching of the gene between an active and an inactive epigenetic states, while the other three reactions represent the synthesis and decay of the protein. The biochemical state of the gene of interest can be described by a pair of variables $(i,n)$, where $i$ represents the gene activity with $i = 1$ and $i = 0$ corresponding to the active and inactive states, respectively, and $n$ stands for the protein copy number.

Let $\alpha(t)$ and $X(t)$ denote the gene activity and protein copy number in a single cell at time $t$, respectively. Let $p_{i,n}(t) = \mathrm{Prob}(\alpha(t)=i, X(t)=n)$ denote the probability of having $n$ copies of proteins at time $t$ when the gene is in state $i$. Then the dynamics of stochastic gene expression can be described by the Delbr\"{u}ck-Gillespie process (DGP) with transition diagram illustrated in Fig. \ref{model}(b). Mathematically, the DGP is a continuous-time Markov chain with infinite state space, whose evolution is governed by the chemical master equation
\begin{equation*}\left\{
\begin{split}
\dot p_{1,n} =&\; a_np_{0,n}+sp_{1,n-1}+(n+1)dp_{1,n+1}\\
&\; -(b_n+s+nd)p_{1,n}, \\
\dot p_{0,n} =&\; b_np_{1,n}+rp_{0,n-1}+(n+1)dp_{0,n+1}\\
&\; -(a_n+r+nd)p_{0,n}.
\end{split}\right.
\end{equation*}
Here $a_n$ and $b_n$ are the switching rates of the gene between the active and inactive states, $s$ and $r$ are the synthesis rates of the protein when the gene is active and inactive, respectively, and $d$ is the decaying rate of the protein either due to protein degradation or due to dilution during cell division \cite{friedman2006linking, ge2015stochastic}. In general, the protein decaying rate can be decomposed as $d = \log 2/T_p+\log 2/T_c$, where $T_p$ is the protein half-life and $T_c$ is the cell cycle time \cite{friedman2006linking}. Because of the effect of autoregulation, the gene switching rates $a_n$ and $b_n$ generally depend on the protein copy number $n$. Since many genes have complex epigenetic controls including dissociation of repressors, association of activators, or chromatin remodeling, we do not impose any restrictions on the specific functional forms of $a_n$ and $b_n$. In the previous studies \cite{hornos2005self, grima2012steady, kumar2014exact, potoyan2015dichotomous, liu2016decomposition}, many authors have studied the case of linear feedback regulation where $a_n$ and $b_n$ depend linearly on $n$. However, recent single-cell experiments on transcription of mammalian cells \cite{bintu2016dynamics} suggest that $a_n$ and $b_n$ are often saturated when $n\gg 1$ and thus are highly nonlinear. In the present work, we consider a more general case by allowing arbitrary nonlinearity.

In most applications, the gene switches rapidly between the active and inactive states, that is, $a_n,b_n\gg s,d$. Let $p_n(t) = p_{0,n}(t)+p_{1,n}(t)$ denote the probability of having $n$ copies of proteins at time $t$. When the gene switching rates are fast, the biochemical states $(1,n)$ and $(0,n)$ can be combined into a single state $n$ \cite{yin2013continuous, jia2016model}. In this way, the original DGP can be simplified to the reduced Markov model illustrated in Fig. \ref{model}(c), whose dynamics is governed by the chemical master equation
\begin{equation*}
\dot p_n = c_{n-1}p_{n-1}+(n+1)dp_{n+1}-(c_n+nd)p_n,
\end{equation*}
where
\begin{equation*}
c_n = \frac{a_ns+b_nr}{a_n+b_n}
\end{equation*}
is the effective synthesis rate of the protein. If the network has no feedback, the gene switching rates $a_n$ and $b_n$ do not depend on the protein copy number $n$ and thus $c_n$ is a constant. In an autoregulatory gene network, the protein binds to its own gene and changes the gene activity. In the case of positive feedback, $c_n$ is an increasing function of $n$, whereas in the case of negative feedback, $c_n$ is a decreasing function of $n$. In realistic biological systems, $c_n$ is usually a nonlinear function and is often described by a generalized Hill function \cite{friedman2006linking, jia2017stochastic}.

Let $\mu_n$ denote the steady-state probability of having $n$ copies of proteins. Since the reduced model satisfies detailed balance, the steady-state probabilities $\mu_n$ can be easily calculated as
\begin{equation}\label{stationary}
\mu_n = \frac{A}{n!d^n}\prod_{k=0}^{n-1}c_k,
\end{equation}
where $A$ is a normalization constant. In general, the dynamics of a gene regulatory network is exponentially ergodic \cite{chen2000equivalence, chen2006eigenvalues, meyn2012markov}, which means that the time-dependent probabilities $p_n(t)$ will converge to the steady-state probabilities $\mu_n$ at exponential speed: for any $n\geq 0$, there exists a constant $K_n>0$ such that
\begin{equation*}
|p_n(t)-\mu_n|\leq K_ne^{-\gamma t},\;\;\;\forall t\geq 0.
\end{equation*}
Here the optimal (maximal) parameter $\gamma$ characterizes the relaxation rate to the steady state. In the mathematics literature, it is called the exponentially ergodic convergence rate. The relaxation time $T$ of the network is defined as the inverse of the relaxation rate, that is, $T = 1/\gamma$.

\section{Spectral gaps for infinite Markov chains}
Given a particular initial state, the evolution of a continuous-time Markov chain is completely determined by its generator matrix. Let us first consider the case when the Markov chain has a finite state space $S = \{0,1,\cdots, N\}$. Let $Q = (q_{mn})$ denote the generator matrix of the Markov chain, where $q_{mn}$ with $m\neq n$ is the transition rate (transition probability per unit time) from state $m$ to state $n$ and $q_{mm} = -\sum_{n\neq m}q_{mn}$. Let $\lambda_0,\cdots,\lambda_N$ denote all the eigenvalues of $Q$. When the Markov chain is irreducible, the Perron-Frobenius theorem states that one eigenvalue of $Q$ must be zero and the real parts of all the other eigenvalues must be negative \cite{berman1979nonnegative}:
\begin{equation*}
0 = \lambda_0 > \textrm{Re}(\lambda_1)\geq \cdots\geq \textrm{Re}(\lambda_N),
\end{equation*}
where $\textrm{Re}(x)$ denotes the real part of $x$. In this case, it is well-known that the relaxation rate $\gamma$ of the Markov chain is exactly $|\textrm{Re}(\lambda_1)|$, the spectral gap between the zero eigenvalue and first nonzero eigenvalue.

The situation becomes much more complicated when the Markov chain has an infinite state space $S = \{0,1,2,\cdots\}$. Recall that any eigenvalue-eigenvector pair $(\lambda,v)$ of the generator matrix $Q$ is related by the characteristic equation $Qv = \lambda v, v\neq 0$. For an infinite Markov chain, if we do not make any restrictions on the eigenvector $v = (v_n)$, then the set of eigenvalues may not be discrete and thus the spectral gap may vanish. To overcome this difficulty, it is useful to introduce the following Banach space:
\begin{equation*}
l^p(\mu) = \left\{v = (v_n): \sum_n|v_n|^p\mu_n<\infty\right\},\;\;\;1\leq p\leq\infty,
\end{equation*}
where $\mu = (\mu_n)$ is the steady-state distribution of the Markov chain. We next introduce a crucial definition. If a vector $v\in l^p(\mu)$ satisfies the characteristic equation $Qv = \lambda v, v\neq 0$, then $\lambda$ is called an eigenvalue in $l^p(\mu)$. In fact, if we require that the eigenvectors $v\in l^p(\mu)$, then the set of eigenvalues may become discrete. Let $\one = (1,1,\cdots)^T$ denote the column vector whose components are all 1. It is easy to see that $0$ must be an eigenvalue in $l^p(\mu)$ because the associated eigenvector $\one\in l^p(\mu)$. The resulting spectral gap in $l^p(\mu)$ will be denoted by $\textrm{Gap}(l^p(\mu))$.

For infinite Markov chains, another strange phenomenon may occur. The spectral gap may vary among different choices of $p$. Fortunately, there is a mathematical theorem \cite{chen1991exponentiall, chen2004markov} which claims that when a Markov chain satisfies detailed balance, the spectral gap in $l^2(\mu)$ with $p = 2$ exactly coincides with the relaxation rate $\gamma$, that is, $\gamma = \textrm{Gap}(l^2(\mu))$. Since the reduced model illustrated in Fig. \ref{model}(c) satisfies detailed balance, the relaxation rate of the network is exactly the spectral gap in $l^2(\mu)$.

We stress here that when $p = 2$, the Banach space $l^2(\mu)$ turns out to be a Hilbert space where an inner product, also called scalar product, can be introduced. The inner product allows the rigorous introduction of intuitive geometrical notions such as the angle between two vectors, especially the orthogonality between two vectors. Furthermore, when a Markov chain satisfies detailed balance, the generator matrix $Q$ happens to be an Hermitian (self-adjoint) operator on the Hilbert space $l^2(\mu)$ and its spectral theory is well-developed in modern probability theory. The above framework is rather similar to that of quantum mechanics, where any physically meaningful observable must be an Hermitian operator on a Hilbert space and its eigenvalues correspond to all possible outcomes of a measurement of the given observable.

\section{Relaxation rates for autoregulatory gene networks}

\subsection{Relaxation rates for networks with no feedback}
Let $Q$ denote the generator matrix of the reduced model. Then the characteristic equation $Qv = \lambda v$ can be written in components as
\begin{equation}\label{characteristic}
nd(v_{n-1}-v_n) + c_n(v_{n+1}-v_n) = \lambda v_n.
\end{equation}
We shall normalize the eigenvector $v=(v_n)$ so that $v_0 = 1$.

As mentioned previously, if the network has no feedback, then $c_n = c$ is a constant. From \eqref{stationary}, the steady-state distribution is given by the Poisson distribution
\begin{equation*}
\mu_n = \frac{(c/d)^n}{n!}e^{-c/d}.
\end{equation*}
To proceed, we define the following power series:
\begin{equation}\label{series}
f(x) = \sum_n\frac{v_n}{n!}x^n.
\end{equation}
Since $v_0 = 1$, we have $f(0) = 1$. Then \eqref{characteristic} can be converted to the following ordinary differential equation:
\begin{equation*}
(c-dx)f'(x)-(c+\lambda-dx)f(x) = 0,\;\;\;f(0) = 1.
\end{equation*}
The solution of this equation is given by
\begin{equation*}
f(x) = e^x\left(1-dx/c\right)^{-\lambda/d}.
\end{equation*}
For convenience, let $(x)_n = x(x+1)\cdots(x+n-1) = \Gamma(x+n)/\Gamma(x)$ denote the Pochhammer symbol. From \eqref{series}, the eigenvector $v = (v_n)$ can be recovered from $f(x)$ as
\begin{equation}\label{eigenvector1}
v_n = f^{(n)}(0) = \sum_{k=0}^nC_n^k(\lambda/d)_k(d/c)^k,
\end{equation}
where $f^{(n)}(x)$ is the $n$th derivative of $f(x)$ and $C_n^k = n!/k!(n-k)!$ is the combinatorial number.

\begin{figure}[!htb]
\begin{center}
\centerline{\includegraphics[width=0.6\textwidth]{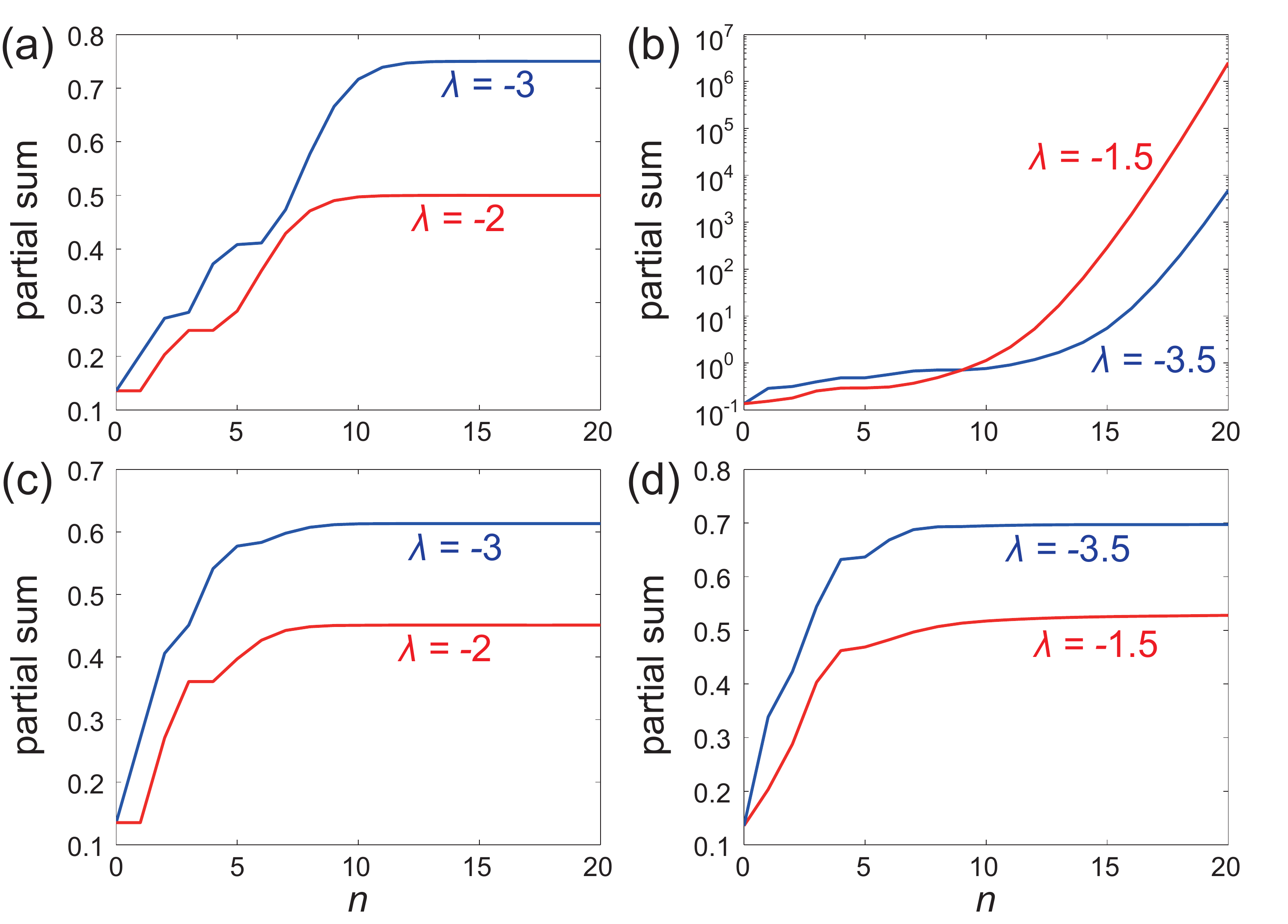}}
\caption{\textbf{Partial sums of four power series under different choices of $\lambda$.} (a) The partial sum of the power series $\sum_n|v_n|^2\mu_n$ for $\lambda = -2$ and $\lambda = -3$. (b) The partial sum of the power series $\sum_n|v_n|^2\mu_n$ for $\lambda = -1.5$ and $\lambda = -3.5$. (c) The partial sum of the power series $\sum_n|v_n|\mu_n$ for $\lambda = -2$ and $\lambda = -3$. (d) The partial sum of the power series $\sum_n|v_n|\mu_n$ for $\lambda = -1.5$ and $\lambda = -3.5$. In (a)-(d), the model parameters are chosen as $c = 2$ and $d = 1$.
}\label{convergence}
\end{center}
\end{figure}

To find all the eigenvalues $\lambda$ in $l^2(\mu)$, we consider two different cases. Let us first focus on the case when $\lambda = -md$ for some integer $m\geq 0$. In this case, the term $(\lambda/d)_k = (-m)_k$ will vanish when $k\geq m+1$. This suggests that $|v_n|\leq 2^n$ when $n\gg 1$ (see Supplementary Material, Sec. 1), which gives rise to
\begin{equation*}
\sum_n|v_n|^2\mu_n \leq e^{-c/d}\sum_n\frac{(4c/d)^n}{n!} = e^{3c/d}<\infty.
\end{equation*}
This shows that $v\in l^2(\mu)$ and thus $\lambda = -md$ is an eigenvalue in $l^2(\mu)$ (see Fig. \ref{convergence}(a)).

Let us next focus on the case when $\lambda\neq -md$ for all integers $m\geq 0$. In this case, $c/d$ is a singular point of $f(x)$ and thus the convergence radius of the power series \eqref{series} is $c/d$. This shows that
\begin{equation*}
\limsup_{n\rightarrow\infty}\sqrt[n]{\frac{|v_n|}{n!}} = d/c.
\end{equation*}
This equation further gives rise to
\begin{equation*}
\limsup_{n\rightarrow\infty}\sqrt[n]{\frac{|v_n|^2}{n!}} = \infty.
\end{equation*}
Thus the convergence radius of the power series $\sum_n|v_n|^2x^n/n!$ is zero, which gives rise to
\begin{equation*}
\sum_n|v_n|^2\mu_n = e^{-c/d}\sum_n\frac{|v_n|^2}{n!}(c/d)^n = \infty.
\end{equation*}
This shows that $v\notin l^2(\mu)$ and thus $\lambda\neq -md$ is not an eigenvalue in $l^2(\mu)$ (see Fig. \ref{convergence}(b)). If the network has no feedback, the above analysis shows that all the eigenvalues in $l^2(\mu)$ are given by
\begin{equation*}
\lambda = 0,-d,-2d,\cdots.
\end{equation*}
Thus the relaxation rate of the network is the spectral gap $d$ in $l^2(\mu)$, which is exactly the decaying rate of the protein.

In fact, the situation becomes completely different if we change $l^2(\mu)$ to $l^1(\mu)$. It can be shown that for any complex number $\lambda$ in the half-left plane, the associated eigenvector $v\in l^1(\mu)$ (see Fig. \ref{convergence}(c),(d)). This implies that the set of all eigenvalues in $l^1(\mu)$ is the whole half-left plane and thus the spectral gap in $l^1(\mu)$ will vanish. Therefore, it is impossible to use the spectral gap in $l^1(\mu)$ to characterize the relaxation rates of gene regulatory networks.

\subsection{Relaxation rates for networks with positive feedback}\label{positive}
In a gene circuit with positive autoregulation, the protein binds and activates its own gene. In this case, $c_n$ is an increasing function of $n$. For simplicity, we assume that $c_n = c+un$ for some $0<u<d$, where $u$ is the strength of positive feedback. Here the condition $u<d$ is the sufficient and necessary condition for the reduced model to have a steady-state distribution. From \eqref{stationary}, the steady-state distribution is given by the negative-binomial distribution
\begin{equation*}
\mu_n = \frac{(u/d)^n}{n!}(c/u)_n(1-u/d)^{-c/u}.
\end{equation*}
Moreover, \eqref{characteristic} can be converted to the following ordinary differential equation:
\begin{equation*}
uxf''(x) + (c-(u+d)x)f'(x) - (c+\lambda-dx)f(x) = 0,\;\;\;f(0) = 1.
\end{equation*}
The solution of this equation is given by
\begin{equation*}
f(x) = e^xM(\alpha,\beta,(d/u-1)x),
\end{equation*}
where $M(\alpha,\beta,x)$ is Kummer's confluent hypergeometric function and
\begin{equation*}
\alpha = \frac{\lambda}{d-u},\;\;\;\beta = \frac{c}{u}.
\end{equation*}
From \eqref{series}, the eigenvector $v = (v_n)$ can be recovered from $f(x)$ as
\begin{equation}\label{eigenvector2}
v_n = f^{(n)}(0) = \sum_{k=0}^nC_n^k\frac{(\alpha)_k}{(\beta)_k}(d/u-1)^k.
\end{equation}

Similarly, we consider two different cases. Let us first focus on the case when $\lambda = -m(d-u)$ for some integer $m\geq 0$. In this case, the term $(\alpha)_k = (-m)_k$ will vanish when $k\geq m+1$. This suggests that for any $\epsilon>0$, we have $|v_n|\leq (1+\epsilon)^n$ when $n\gg 1$ (see Supplementary Material, Sec. 2). This inequality gives rise to
\begin{equation*}
\limsup_{n\rightarrow\infty}\sqrt[n]{|v_n|} \leq 1.
\end{equation*}
By the Stirling formula for the gamma function, it is easy to prove that (see Supplementary Material, Sec. 3)
\begin{equation}\label{stirling}
\lim_{n\rightarrow\infty}\sqrt[n]{\frac{(\beta)_n}{n!}} = 1.
\end{equation}
The above two equations indicate that
\begin{equation*}
\limsup_{n\rightarrow\infty}\sqrt[n]{\frac{|v_n|^2(\beta)_n}{n!}} \leq 1.
\end{equation*}
Thus the convergence radius of the power series $\sum_n|v_n|^2(\beta)_nx^n/n!$ is larger than or equal to 1, which gives rise to
\begin{equation*}
\sum_n|v_n|^2\mu_n = (1-u/d)^{-c/u}\sum_n\frac{|v_n|^2(\beta)_n}{n!}(u/d)^n < \infty.
\end{equation*}
This shows that $v\in l^2(\mu)$ and thus $\lambda = -m(d-u)$ is an eigenvalue in $l^2(\mu)$.

Let us next focus on the case when $\lambda\neq -m(d-u)$ for all integers $m\geq 0$. In this case, it can be proved that (see Supplementary Material, Sec. 3)
\begin{equation*}
\limsup_{n\rightarrow\infty}\sqrt[n]{|v_n|} \geq d/u.
\end{equation*}
This equation, together with \eqref{stirling}, shows that
\begin{equation*}
\limsup_{n\rightarrow\infty}\sqrt[n]{\frac{|v_n|^2(\beta)_n}{n!}} \geq (d/u)^2.
\end{equation*}
Thus the convergence radius of the power series $\sum_{n}|v_n|^2(\beta)_nx^n/n!$ is smaller than or equal to $(u/d)^2$, which gives rise to
\begin{equation*}
\sum_n|v_n|^2\mu_n = (1-u/d)^{-c/u}\sum_n\frac{|v_n|^2(\beta)_n}{n!}(u/d)^n = \infty.
\end{equation*}
This shows that $v\notin l^2(\mu)$ and thus $\lambda\neq -m(d-u)$ is not an eigenvalue in $l^2(\mu)$. In the positive-feedback case, the above analysis shows that all the eigenvalues in $l^2(\mu)$ are given by
\begin{equation*}
\lambda = 0,-(d-u),-2(d-u),\cdots.
\end{equation*}
Thus the relaxation rate of the network is the spectral gap $d-u$ in $l^2(\mu)$, which is smaller than the decaying rate of the protein.

The readers may ask what will happen if $c_n$ does not have the form of $c_n = c+un$. In fact, it can be proved that whenever $c_n$ is an increasing function of $n$, the relaxation rate of the network is always smaller than the decaying rate of the protein (see Supplementary Material, Sec. 4). This shows that positive feedback will always slow down the relaxation to the steady state.

\subsection{Relaxation rates for networks with negative feedback}
In a gene circuit with negative autoregulation, the protein binds and inhibits its own gene. In this case, $c_n$ is a decreasing function of $n$ and it is almost impossible to obtain an explicit expression of the eigenvalues in $l^2(\mu)$. Fortunately, there is a mathematical theorem which claims that the spectral gap of the reduced model in $l^2(\mu)$ has the following variational formula \cite{chen1996estimation}:
\begin{equation*}
\textrm{Gap}(l^2(\mu)) = \sup_{v>0}\inf_{n\geq 0}\{(n+1)d+c_n-\frac{nd}{v_{n-1}}-c_{n+1}v_n\},
\end{equation*}
where $v>0$ means that $v = (v_n)$ ranges over all vectors such that $v_n>0$ for each $n$ and $v_{-1}$ is defined as $v_{-1} = 1$. Taking $v_n = 1$ for each $n$ in the above equation gives rise to
\begin{equation*}
\textrm{Gap}(l^2(\mu)) \geq \inf_{n\geq 0}\{d+c_n-c_{n+1}\}.
\end{equation*}
Since $c_n$ is a decreasing function of $n$, we have $c_n-c_{n+1}\geq 0$ for each $n$. This finally shows that
\begin{equation}\label{inequality}
\textrm{Gap}(l^2(\mu)) \geq d.
\end{equation}
Thus the relaxation rate of the network is larger than the decaying rate of the protein. This shows that negative feedback will always speed up the relaxation to the steady state.

\section{Numerical simulations}

\subsection{Inference of the feedback topology}
Our theoretical results provide a novel possible method to infer the feedback topology of an autoregulatory gene network based on (single-cell or bulk) time-series data of gene expression. Suppose that in an autoregulatory circuit, the expression levels of the gene within a population of cells have been measured at several discrete times. Then the relaxation rate $\gamma$ can be estimated as the exponentially decaying rate of the ensemble average $\langle A_{X(t)}\rangle$ associated with some appropriate observable $A = (A_n)$, where $X(t)$ denotes the protein abundance (copy number, concentration, or fluorescence) at time $t$. Specifically, when $t\gg 1$, the ensemble average of the observable $A$ at time $t$ has the following approximation:
\begin{equation*}
\langle A_{X(t)}\rangle \approx \langle A_{X(\infty)}\rangle+Ke^{-\gamma t},
\end{equation*}
where $\langle A_{X(\infty)}\rangle = \sum_nA_n\mu_n$ is the ensemble average at the steady state and $K$ is a constant. Taking logarithm on both sides of this equation gives rise to
\begin{equation*}
\log|\langle A_{X(t)}\rangle-\langle A_{X(\infty)}\rangle| \approx \log|K|-\gamma t,
\end{equation*}
which is a linear relation with respect to time $t$. We only need to calculate the logarithm of the difference between the time-dependent observation and the steady-state observation and then carry out a linear regression analysis with respect to time $t$ when $t\gg 1$. The slope of the linear regression provides a robust estimation of the relaxation rate $\gamma$.

Recent experiments have also made it possible to measure the decaying rate $d$ of the corresponding protein \cite{belle2006quantification, eden2011proteome, moran2013sizing, christiano2014global}. By comparing the two numbers $\gamma$ and $d$, we may infer the feedback topology of the underlying autoregulatory network. If $\gamma$ is significantly larger (smaller) than $d$, we have good reasons to believe that there is a negative-feedback (positive-feedback) loop regulating this gene.
\begin{figure}[!htb]
\centerline{\includegraphics[width=0.6\textwidth]{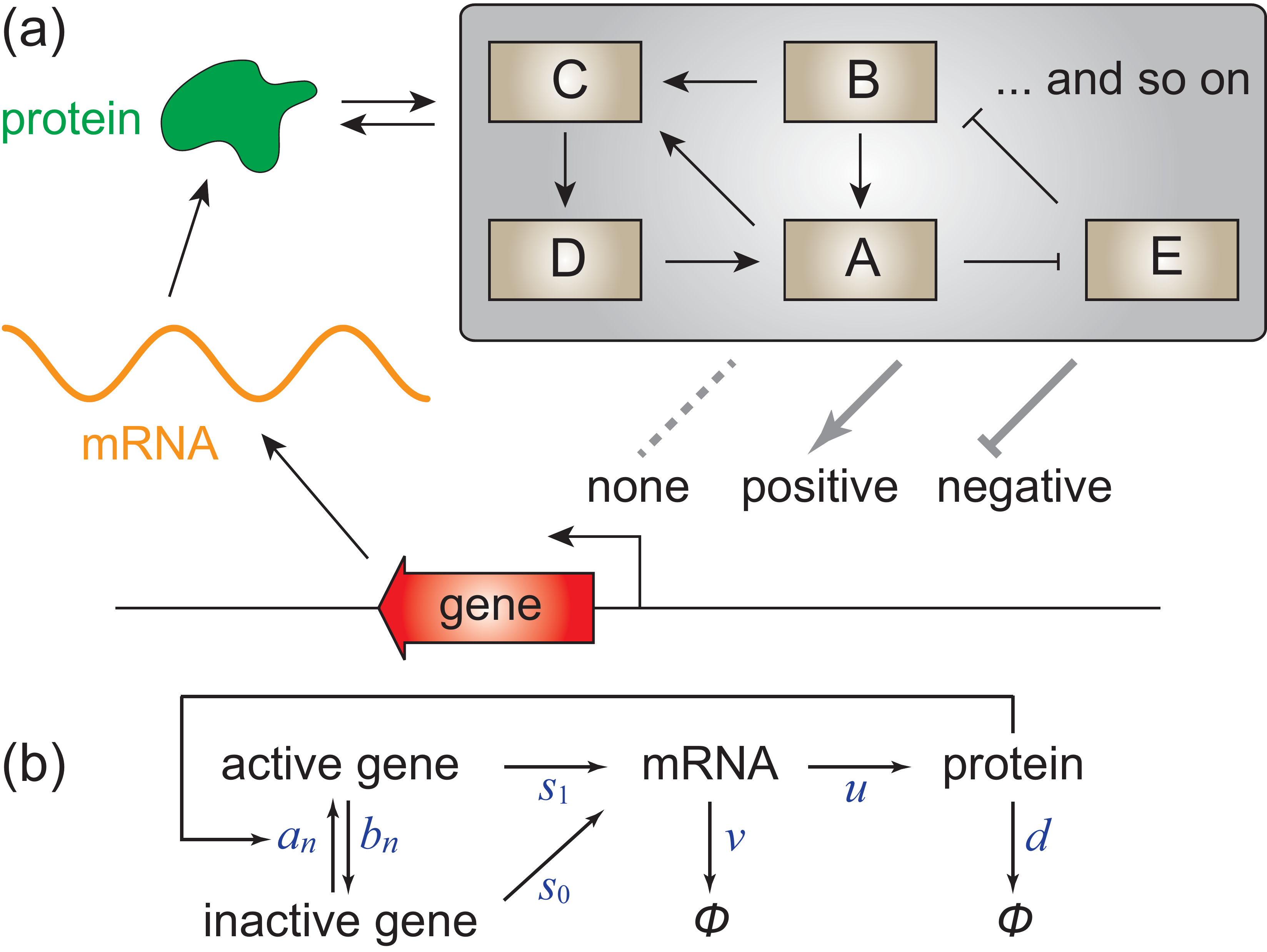}}
\caption{\textbf{Models of stochastic gene expression with the mRNA kinetics.} (a) Gene regulatory networks in a living cell can be overwhelmingly complex, involving numerous feedback loops and signaling steps. However, if one focuses on a particular gene of interest (red), then there are only three types of fundamental regulatory relations: no feedback (none), positive feedback, and negative feedback. The dotted line denotes that there is no link between adjacent nodes. (b) The three-stage model of stochastic gene expression.}\label{network}
\end{figure}

A natural question is whether our theory can be applied to more complex gene regulatory networks. In general, if we focus on a particular gene of interest and the feedback loop regulating this gene, there are three types of gross topological structures: no feedback, positive feedback, and negative feedback (see Fig. \ref{network}(a)). Intuitively, the relaxation of a complex gene network may be delayed by a long feedback loop, which limits the application of the above method. However, there is a special case where our theory is still applicable. If the protein decaying rate of the target gene is much slower than those of the other genes, then the target gene is a slow variable and the other genes are fast variables. By using the classical simplification techniques of two-time-scale Markov chains \cite{yin2013continuous, jia2016model}, the fast variables can be coarse-grained and thus the complex network can be reduced to a much simpler network which only involves autoregulation of the target gene. This implies that our theory can only be applied to infer the gross feedback topology of the target gene with the slowest protein decaying rate. More generally, for a target gene with a relatively fast protein decaying rate, if a biological experiment can be designed to reduce its protein decaying rate so that it becomes much slower than those of the other genes, then our theory might be applied to infer the gross feedback topology of this gene.

\subsection{Effects of translation bursts on the relaxation rate}
We would like to point out that the gene expression model illustrated in Fig. \ref{model}(a) ignores the mRNA kinetics. A more complete model of stochastic gene expression is the three-stage representation involving gene switching, transcription, and translation, as illustrated in Fig. \ref{network}(b) \cite{shahrezaei2008analytical}. Experimentally, it has been consistently observed that the mRNA decays substantially faster than its protein counterpart \cite{shahrezaei2008analytical, moran2013sizing, jia2017simplification}, in which case protein synthesis will occur in random bursts \cite{friedman2006linking, jia2017simplification}. A natural question is whether the main results of this paper still hold when the mRNA kinetics and translation bursts are taken into account.

\begin{figure*}[!htb]
\begin{center}
\centerline{\includegraphics[width=1.0\textwidth]{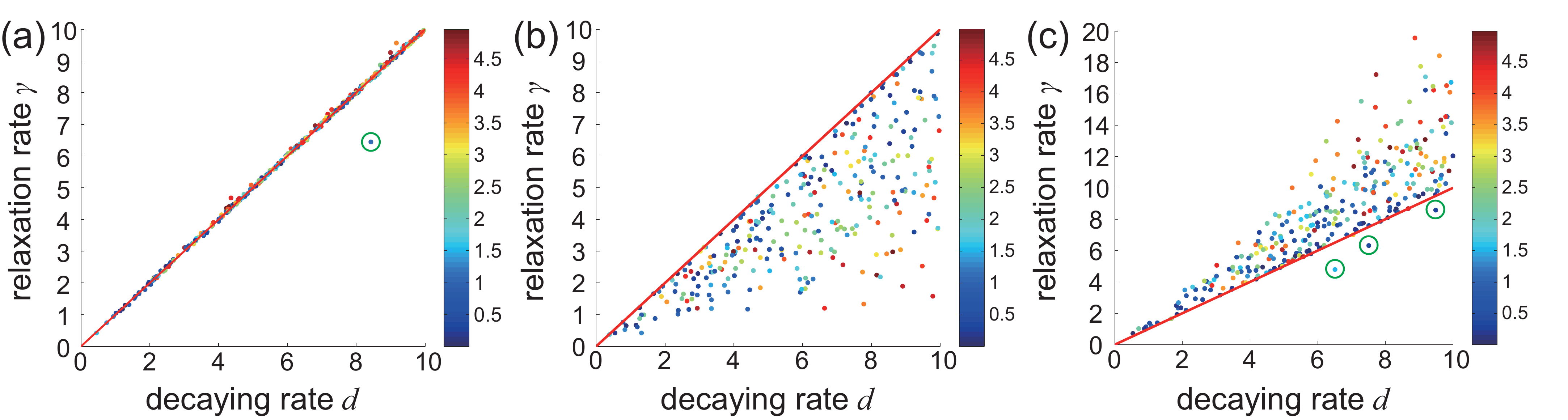}}
\caption{\textbf{Numerical simulations of the relaxation rate of the complete three-stage model under different choices of model parameters.} (a) Networks with no feedback. (b) Networks with positive feedback. (c) Networks with negative feedback. In (a)-(c), the red line represents the line of $\gamma = d$ and the model parameters are randomly chosen as $r\sim U[0,10], s\sim U[r,100], d\sim U[0,10], v = 30d, u/v\sim U[0,5]$, where $U[x,y]$ denotes the uniform distribution over the interval $[x,y]$. The colors of different sample points represent the values of the mean burst size $u/v$ of the protein, as demonstrated by the associated colorbars. The functional forms of the gene switching rates are chosen as $a_n = a, b_n = b$ with $a,b\sim U[0,100]$ in (a), $a_n = a+un, b_n = b$ with $a,u\sim U[0,10], b\sim U[0,100]$ in (b), and $a_n = a, b_n = b+un$ with $a\sim U[0,100], b,u\sim U[0,10]$ in (c).}\label{simulation}
\end{center}
\end{figure*}

The three-stage model does not satisfy detailed balance, even when the gene switching rates are very fast. Therefore, the existing mathematical tools may not be enough to present a rigorous mathematical analysis on its relaxation kinetics. In Fig. \ref{simulation}(a)-(c), using the Gillespie algorithm, we numerically simulate the relaxation rate of the complete three-stage model in three types of gene networks under different choices of model parameters. Interestingly, according to our simulations, the main results of this paper are still valid in the presence of translation bursts. In the present work, all the derivations are based on the assumption of fast gene switching. However, it can be seen that the simulation results coincide with our theoretical results reasonably well, even when the gene switching rates are relatively small. In networks with no feedback and negative feedback, there is a very small portion of outliers lying on the bottom of the red line (see the green circles in Fig. \ref{simulation}(a),(c)). We examine these outliers in detail and find that the gene switching rates of them are extremely small. This suggests that extremely small gene switching rates may slow down the relaxation to the steady state. In positive-feedback networks, the outliers are not detected and the relaxation rate is always smaller than the protein decaying rate.

For networks with linear or nonlinear feedback regulation, the relaxation rate no longer coincides with the protein decaying rate. Interestingly, we find that the deviation of the relaxation rate from the protein decaying rate is remarkably affected by the mean burst size $u/v$ of the protein, which is defined as the averaged number of proteins produced by each mRNA molecule before it is degraded \cite{jia2017simplification, jia2017emergent, jia2017stochastic}. In Fig. \ref{simulation}(a)-(c), we paint all sample points in different colors representing the values of the protein burst size $u/v$, as demonstrated by the associated colorbars. As can be seen from Fig. \ref{simulation}(b),(c), there is an remarkable trend that networks with larger protein burst sizes tend to possess slower relaxation rates in the positive-feedback case and tend to possess faster relaxation rates in the negative-feedback case. In other words, for autoregulatory gene networks, an increase in the protein burst size tends to enlarge the deviation of the relaxation rate from the protein decaying rate.

\subsection{Effects of the feedback strength on the relaxation rate}
\begin{figure}[!htb]
\centerline{\includegraphics[width=0.4\textwidth]{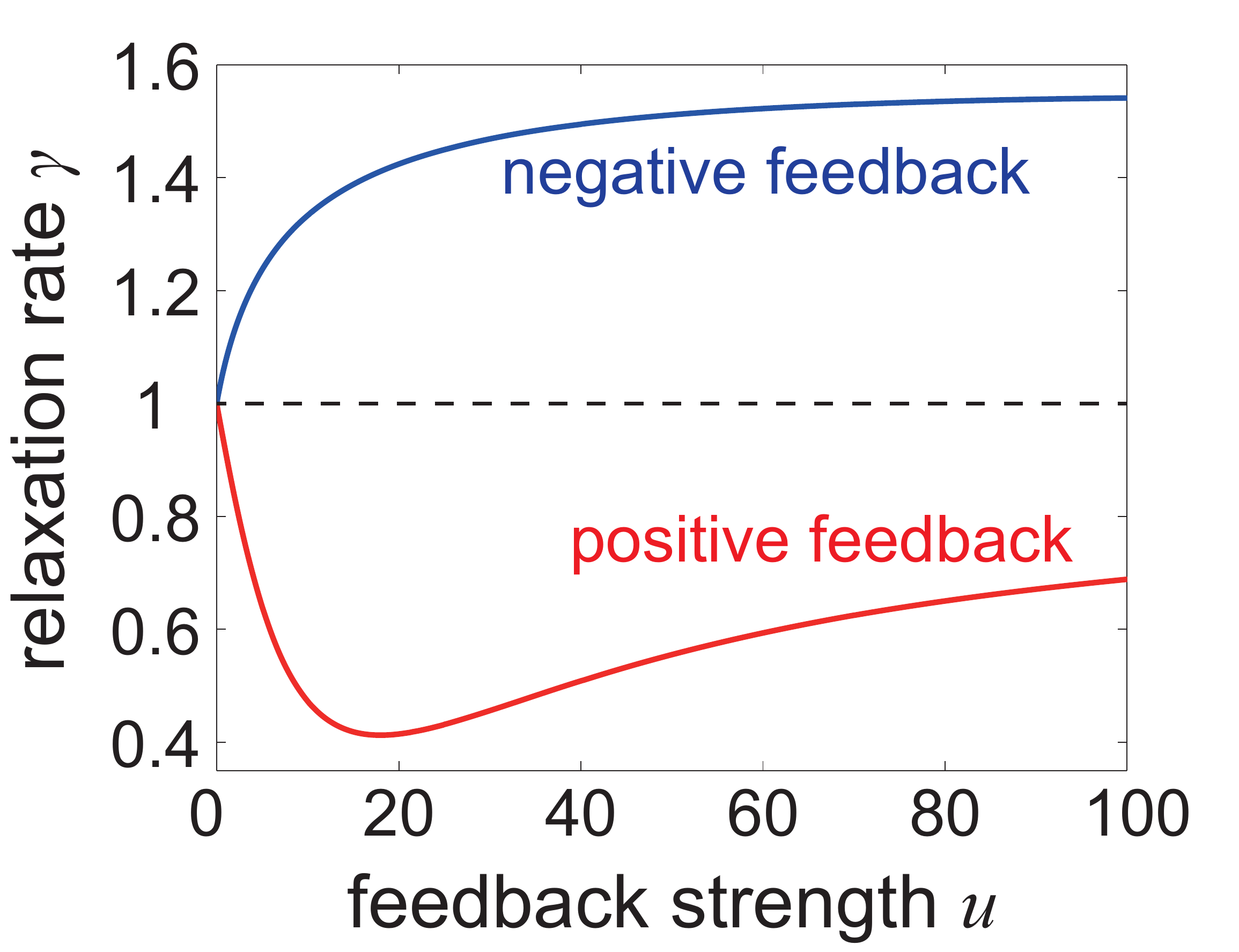}}
\caption{\textbf{The relaxation rate $\gamma$ versus the feedback strength $u$ in networks with positive (red) and negative (blue) autoregulation.} The model parameters are chosen as $s = 10, r = 1, d = 1$. The gene switching rates are chosen as $a = 0, b = 100$ in the positive-feedback case and $a = 100, b = 0$ in the negative-feedback case.}\label{comparison}
\end{figure}

To further clarify the effects of feedback regulation on the relaxation kinetics, we numerically simulate the relaxation rates of autoregulatory gene networks under a set of biologically relevant parameters. We first consider networks with positive autoregulation. For simplicity, we assume that the gene switching rates have the form of $a_n = a+un$ and $b_n = b$ \cite{kumar2014exact}, where $a$ is the spontaneous switching rate from the inactive state to the active state and $u$ is the strength of positive feedback. In this case, the effective synthesis rate of the protein is given by
\begin{equation*}
c_n = \frac{as+br+usn}{a+b+un}.
\end{equation*}
Compared with the linear relation $c_n = c+un$ discussed in Sec. \ref{positive}, this effective protein synthesis rate is saturated when $n\gg 1$ and corresponds to a more realistic situation. Fig. \ref{comparison} illustrates the relaxation rate $\gamma$ as a function of the feedback strength $u$, from which we see that the relaxation rate $\gamma$ is always smaller than the protein decaying rate $d$ and tends to the protein decaying rate $d$ as positive feedback becomes increasingly weaker. Moreover, the relaxation rate is also saturated when the feedback strength $u\gg 1$, which is different from the linear case discussed in Sec. \ref{positive}.

We next focus on networks with negative autoregulation. For simplicity, we assume that the gene switching rates have the form of $a_n = a$ and $b_n = b+un$ with $u$ being the strength of negative feedback. As illustrated in Fig. \ref{comparison}, the relaxation rate $\gamma$ is always larger than the protein decaying rate $d$ and the equal sign in \eqref{inequality} is reached as the negative-feedback network evolves to a network with no feedback, that is, $u\rightarrow 0$. There is a crucial difference between networks with positive and negative feedback. In the negative-feedback case, the relaxation rate is a monotonic function of the feedback strength. However, in the positive-feedback case, as positive feedback becomes stronger, the relaxation rate first reaches a low valley, and then increases to the saturated value.

\section{Discussion and conclusions}
In this paper, we present a rigorous mathematical analysis on the relaxation kinetics of autoregulatory gene networks based on the chemical master equation model of single-cell stochastic gene expression. It is worth noting that our results depend nothing on the specific functional form of the effective protein synthesis rate $c_n$ except for its monotonicity, which makes our theory highly general. In the previous studies, many authors have studied the case of linear feedback regulation where $a_n$ and $b_n$ depend linearly on $n$. One of the most powerful parts of our theory is that it can be applied to gene regulatory networks with nonlinear feedback regulation. If a network has no feedback, the relaxation rate to the steady state is exactly the decaying rate of the protein. Furthermore, we show that positive feedback always slows down the relaxation kinetics, while negative feedback always speeds it up.

We stress that in this paper, we focus on the recovery of a cell to the steady state in the long-time regime. Our analytical approach can not be applied to study the response behavior to an external perturbation in the short-time regime. In fact, the response behavior of autoregulatory gene networks at small times has been studied in the previous literature based on deterministic models \cite{rosenfeld2002negative}. Moreover, some authors have also established the fluctuation-dissipation theorems for stochastic Markov systems \cite{chen2017fluctuation, chen2017mathematical}, which can be used to study the linear response behavior of gene regulatory models \cite{yan2013fluctuation} and some sensory systems such as bacterial chemotaxis \cite{jia2017nonequilibrium}.

There is another difference between finite and infinite Markov chains. For finite Markov chains, the eigenvalues for $Q$ and $Q^T$ are exactly the same, where $Q^T$ is the transpose of $Q$. However, this is not the case for infinite Markov chains. In \cite{assaf2006spectral}, the authors have established the spectral theory of $Q^T$ for a branching-annihilation chemical reaction with the aid of the Sturm-Liouville theory. In probability theory, however, there is hardly any general relationship between the spectral gap of $Q^T$ and the relaxation rate. This is why we develop the spectral theory of $Q$ in this paper, instead of $Q^T$, for the DGP kinetics of stochastic gene expression.

Rigorous mathematical investigations of stochastic gene expression kinetics has become increasingly important for quantitatively analyzing and understanding molecular complexity of gene regulatory networks at the single-cell and single-molecule levels \cite{hilfinger2016exploiting, estrada2016information}. Here we have presented a theory which links the relaxation time to the feedback topology of the underlying gene circuit. Our results provide a theoretical foundation for designing new kinds of single-cell time-series experiments. This follows the same spirit of what Rob Phillips would call the ``Figure 7 theory" \cite{phillips2015theory}. We anticipate that rapid advances of high-throughput and high-resolution experimental technologies will soon be able to test and sharpen our theory.

\section*{Supplementary material}
See Supplementary Material for the detailed derivations of some nontrivial mathematical formulas and results.

\section*{Acknowledgements}
We are grateful to the anonymous referees for their valuable comments and suggestions which helped us greatly in improving the quality of this paper. H. Qian was supported by NIH Grant R01GM109964. M. Chen was supported by NIH Grant K25AR063761. M.Q. Zhang was supported by NIH Grants MH102616, MH109665 and also by NSFC Grants 31671384 and 91329000.

%%%%%%%%%% 参考文献 %%%%%%%%%%
\setlength{\bibsep}{5pt}
\small\bibliographystyle{nature}

\begin{thebibliography}{66}
\expandafter\ifx\csname natexlab\endcsname\relax\def\natexlab#1{#1}\fi
\expandafter\ifx\csname url\endcsname\relax
  \def\url#1{\texttt{#1}}\fi
\expandafter\ifx\csname urlprefix\endcsname\relax\def\urlprefix{URL }\fi

\bibitem[{McAdams \& Arkin(1997)}]{mcadams1997stochastic}
McAdams, H.~H. \& Arkin, A.
\newblock Stochastic mechanisms in gene expression.
\newblock \emph{Proc. Natl. Acad. Sci. USA} \textbf{94}, 814--819 (1997).

\bibitem[{Elowitz \emph{et~al.}(2002)Elowitz, Levine, Siggia \&
  Swain}]{elowitz2002stochastic}
Elowitz, M.~B., Levine, A.~J., Siggia, E.~D. \& Swain, P.~S.
\newblock Stochastic gene expression in a single cell.
\newblock \emph{Science} \textbf{297}, 1183 (2002).

\bibitem[{Ozbudak \emph{et~al.}(2002)Ozbudak, Thattai, Kurtser, Grossman \& van
  Oudenaarden}]{ozbudak2002regulation}
Ozbudak, E.~M., Thattai, M., Kurtser, I., Grossman, A.~D. \& van Oudenaarden,
  A.
\newblock Regulation of noise in the expression of a single gene.
\newblock \emph{Nat. Genet.} \textbf{31}, 69--73 (2002).

\bibitem[{Blake \emph{et~al.}(2003)Blake, K{\ae}rn, Cantor \&
  Collins}]{blake2003noise}
Blake, W.~J., K{\ae}rn, M., Cantor, C.~R. \& Collins, J.~J.
\newblock Noise in eukaryotic gene expression.
\newblock \emph{Nature} \textbf{422}, 633--637 (2003).

\bibitem[{Raser \& O'Shea(2004)}]{raser2004control}
Raser, J.~M. \& O'Shea, E.~K.
\newblock Control of stochasticity in eukaryotic gene expression.
\newblock \emph{Science} \textbf{304}, 1811--1814 (2004).

\bibitem[{K{\ae}rn \emph{et~al.}(2005)K{\ae}rn, Elston, Blake \&
  Collins}]{kaern2005stochasticity}
K{\ae}rn, M., Elston, T.~C., Blake, W.~J. \& Collins, J.~J.
\newblock Stochasticity in gene expression: from theories to phenotypes.
\newblock \emph{Nat. Rev. Genet.} \textbf{6}, 451--464 (2005).

\bibitem[{Raser \& O'Shea(2005)}]{raser2005noise}
Raser, J.~M. \& O'Shea, E.~K.
\newblock Noise in gene expression: origins, consequences, and control.
\newblock \emph{Science} \textbf{309}, 2010--2013 (2005).

\bibitem[{Golding \emph{et~al.}(2005)Golding, Paulsson, Zawilski \&
  Cox}]{golding2005real}
Golding, I., Paulsson, J., Zawilski, S.~M. \& Cox, E.~C.
\newblock Real-time kinetics of gene activity in individual bacteria.
\newblock \emph{Cell} \textbf{123}, 1025--1036 (2005).

\bibitem[{Raj \emph{et~al.}(2006)Raj, Peskin, Tranchina, Vargas \&
  Tyagi}]{raj2006stochastic}
Raj, A., Peskin, C.~S., Tranchina, D., Vargas, D.~Y. \& Tyagi, S.
\newblock {Stochastic mRNA synthesis in mammalian cells}.
\newblock \emph{PLoS Biol.} \textbf{4}, e309 (2006).

\bibitem[{Xie \emph{et~al.}(2008)Xie, Choi, Li, Lee \& Lia}]{xie2008single}
Xie, X.~S., Choi, P.~J., Li, G.-W., Lee, N.~K. \& Lia, G.
\newblock Single-molecule approach to molecular biology in living bacterial
  cells.
\newblock \emph{Annu. Rev. Biophys.} \textbf{37}, 417--444 (2008).

\bibitem[{Leontovich(1935)}]{leontovich1935basic}
Leontovich, M.~A.
\newblock Basic equations of kinetic gas theory from the viewpoint of the
  theory of random processes.
\newblock \emph{J. Exp. Theoret. Phys.} \textbf{5}, 211--231 (1935).

\bibitem[{Delbr\"{u}ck(1940)}]{delbruck1940statistical}
Delbr\"{u}ck, M.
\newblock Statistical fluctuations in autocatalytic reactions.
\newblock \emph{J. Chem. Phys.} \textbf{8}, 120--124 (1940).

\bibitem[{Anderson \& Kurtz(2015)}]{anderson2015stochastic}
Anderson, D.~F. \& Kurtz, T.~G.
\newblock \emph{{Stochastic Analysis of Biochemical Systems}} (Springer, 2015).

\bibitem[{Milo \emph{et~al.}(2002)}]{milo2002network}
Milo, R. \emph{et~al.}
\newblock Network motifs: simple building blocks of complex networks.
\newblock \emph{Science} \textbf{298}, 824--827 (2002).

\bibitem[{Hilfinger \emph{et~al.}(2016)Hilfinger, Norman \&
  Paulsson}]{hilfinger2016exploiting}
Hilfinger, A., Norman, T.~M. \& Paulsson, J.
\newblock Exploiting natural fluctuations to identify kinetic mechanisms in
  sparsely characterized systems.
\newblock \emph{Cell Systems} \textbf{2}, 251--259 (2016).

\bibitem[{Paulsson \emph{et~al.}(2000)Paulsson, Berg \&
  Ehrenberg}]{paulsson2000stochastic}
Paulsson, J., Berg, O.~G. \& Ehrenberg, M.
\newblock Stochastic focusing: fluctuation-enhanced sensitivity of
  intracellular regulation.
\newblock \emph{Proc. Natl. Acad. Sci. USA} \textbf{97}, 7148--7153 (2000).

\bibitem[{Paulsson \& Ehrenberg(2000)}]{paulsson2000random}
Paulsson, J. \& Ehrenberg, M.
\newblock Random signal fluctuations can reduce random fluctuations in
  regulated components of chemical regulatory networks.
\newblock \emph{Phys. Rev. Lett.} \textbf{84}, 5447 (2000).

\bibitem[{Kepler \& Elston(2001)}]{kepler2001stochasticity}
Kepler, T.~B. \& Elston, T.~C.
\newblock Stochasticity in transcriptional regulation: origins, consequences,
  and mathematical representations.
\newblock \emph{Biophys. J.} \textbf{81}, 3116--3136 (2001).

\bibitem[{Karmakar \& Bose(2004)}]{karmakar2004graded}
Karmakar, R. \& Bose, I.
\newblock Graded and binary responses in stochastic gene expression.
\newblock \emph{Phys. Biol.} \textbf{1}, 197 (2004).

\bibitem[{Walczak \emph{et~al.}(2005)Walczak, Onuchic \&
  Wolynes}]{walczak2005absolute}
Walczak, A.~M., Onuchic, J.~N. \& Wolynes, P.~G.
\newblock Absolute rate theories of epigenetic stability.
\newblock \emph{Proc. Natl. Acad. Sci. USA} \textbf{102}, 18926--18931 (2005).

\bibitem[{Friedman \emph{et~al.}(2006)Friedman, Cai \&
  Xie}]{friedman2006linking}
Friedman, N., Cai, L. \& Xie, X.~S.
\newblock Linking stochastic dynamics to population distribution: an analytical
  framework of gene expression.
\newblock \emph{Phys. Rev. Lett.} \textbf{97}, 168302 (2006).

\bibitem[{Shahrezaei \& Swain(2008)}]{shahrezaei2008analytical}
Shahrezaei, V. \& Swain, P.~S.
\newblock Analytical distributions for stochastic gene expression.
\newblock \emph{Proc. Natl. Acad. Sci. USA} \textbf{105}, 17256--17261 (2008).

\bibitem[{Taniguchi \emph{et~al.}(2010)}]{taniguchi2010quantifying}
Taniguchi, Y. \emph{et~al.}
\newblock {Quantifying E. coli proteome and transcriptome with single-molecule
  sensitivity in single cells}.
\newblock \emph{Science} \textbf{329}, 533--538 (2010).

\bibitem[{Assaf \emph{et~al.}(2011)Assaf, Roberts \&
  Luthey-Schulten}]{assaf2011determining}
Assaf, M., Roberts, E. \& Luthey-Schulten, Z.
\newblock {Determining the stability of genetic switches: explicitly accounting
  for mRNA noise}.
\newblock \emph{Phys. Rev. Lett.} \textbf{106}, 248102 (2011).

\bibitem[{Kumar \emph{et~al.}(2014)Kumar, Platini \& Kulkarni}]{kumar2014exact}
Kumar, N., Platini, T. \& Kulkarni, R.~V.
\newblock Exact distributions for stochastic gene expression models with
  bursting and feedback.
\newblock \emph{Phys. Rev. Lett.} \textbf{113}, 268105 (2014).

\bibitem[{Petrosyan \& Hu(2014)}]{petrosyan2014oscillations}
Petrosyan, K. \& Hu, C.-K.
\newblock Oscillations in probability distributions for stochastic gene
  expression.
\newblock \emph{J. Chem. Phys.} \textbf{140}, 05B623\_1 (2014).

\bibitem[{Jia \emph{et~al.}(2014)Jia, Qian, Kang \& Jiang}]{jia2014modeling}
Jia, C., Qian, M., Kang, Y. \& Jiang, D.
\newblock Modeling stochastic phenotype switching and bet-hedging in bacteria:
  stochastic nonlinear dynamics and critical state identification.
\newblock \emph{Quant. Biol.} \textbf{2}, 110--125 (2014).

\bibitem[{P{\'a}jaro \emph{et~al.}(2015)P{\'a}jaro, Alonso \&
  V{\'a}zquez}]{pajaro2015shaping}
P{\'a}jaro, M., Alonso, A.~A. \& V{\'a}zquez, C.
\newblock Shaping protein distributions in stochastic self-regulated gene
  expression networks.
\newblock \emph{Phys. Rev. E} \textbf{92}, 032712 (2015).

\bibitem[{Lin \& Doering(2016)}]{lin2016gene}
Lin, Y.~T. \& Doering, C.~R.
\newblock Gene expression dynamics with stochastic bursts: Construction and
  exact results for a coarse-grained model.
\newblock \emph{Phys. Rev. E} \textbf{93}, 022409 (2016).

\bibitem[{P{\'a}jaro \emph{et~al.}(2017)P{\'a}jaro, Alonso, Otero-Muras \&
  V{\'a}zquez}]{pajaro2017stochastic}
P{\'a}jaro, M., Alonso, A.~A., Otero-Muras, I. \& V{\'a}zquez, C.
\newblock Stochastic modeling and numerical simulation of gene regulatory
  networks with protein bursting.
\newblock \emph{Journal of Theoretical Biology} \textbf{421}, 51--70 (2017).

\bibitem[{Bressloff(2017)}]{bressloff2017stochastic}
Bressloff, P.~C.
\newblock Stochastic switching in biology: from genotype to phenotype.
\newblock \emph{J. Phys. A: Math. Theor.} \textbf{50}, 133001 (2017).

\bibitem[{Jia(2017{\natexlab{a}})}]{jia2017simplification}
Jia, C.
\newblock Simplification of Markov chains with infinite state space and the
  mathematical theory of random gene expression bursts.
\newblock \emph{Phys. Rev. E} \textbf{96}, 032402 (2017{\natexlab{a}}).

\bibitem[{Jia \emph{et~al.}(2017{\natexlab{a}})Jia, Zhang \&
  Hong}]{jia2017emergent}
Jia, C., Zhang, M.~Q. \& Hong, Q.
\newblock Emergent Lévy behavior in single-cell stochastic gene expression.
\newblock \emph{Phys. Rev. E} \textbf{96}, 040402(R) (2017{\natexlab{a}}).

\bibitem[{Becskei \emph{et~al.}(2001)Becskei, S{\'e}raphin \&
  Serrano}]{becskei2001positive}
Becskei, A., S{\'e}raphin, B. \& Serrano, L.
\newblock Positive feedback in eukaryotic gene networks: cell differentiation
  by graded to binary response conversion.
\newblock \emph{The EMBO Journal} \textbf{20}, 2528--2535 (2001).

\bibitem[{Becskei \& Serrano(2000)}]{becskei2000engineering}
Becskei, A. \& Serrano, L.
\newblock Engineering stability in gene networks by autoregulation.
\newblock \emph{Nature} \textbf{405}, 590--593 (2000).

\bibitem[{Shimoga \emph{et~al.}(2013)Shimoga, White, Li, Sontag \&
  Bleris}]{shimoga2013synthetic}
Shimoga, V., White, J.~T., Li, Y., Sontag, E. \& Bleris, L.
\newblock Synthetic mammalian transgene negative autoregulation.
\newblock \emph{Mol. Syst. Biol.} \textbf{9}, 670 (2013).

\bibitem[{Jia \emph{et~al.}(2017{\natexlab{b}})Jia, Xie, Chen \&
  Zhang}]{jia2017stochastic}
Jia, C., Xie, P., Chen, M. \& Zhang, M.~Q.
\newblock Stochastic fluctuations can reveal the feedback signs of gene
  regulatory networks at the single-molecule level.
\newblock \emph{Sci. Rep.} \textbf{7}, 16037 (2017{\natexlab{b}}).

\bibitem[{Rosenfeld \emph{et~al.}(2002)Rosenfeld, Elowitz \&
  Alon}]{rosenfeld2002negative}
Rosenfeld, N., Elowitz, M.~B. \& Alon, U.
\newblock Negative autoregulation speeds the response times of transcription
  networks.
\newblock \emph{J. Mol. Biol.} \textbf{323}, 785--793 (2002).

\bibitem[{Belle \emph{et~al.}(2006)Belle, Tanay, Bitincka, Shamir \&
  O’Shea}]{belle2006quantification}
Belle, A., Tanay, A., Bitincka, L., Shamir, R. \& O’Shea, E.~K.
\newblock Quantification of protein half-lives in the budding yeast proteome.
\newblock \emph{Proceedings of the National Academy of Sciences} \textbf{103},
  13004--13009 (2006).

\bibitem[{Eden \emph{et~al.}(2011)}]{eden2011proteome}
Eden, E. \emph{et~al.}
\newblock Proteome half-life dynamics in living human cells.
\newblock \emph{Science} \textbf{331}, 764--768 (2011).

\bibitem[{Moran \emph{et~al.}(2013)}]{moran2013sizing}
Moran, M.~A. \emph{et~al.}
\newblock Sizing up metatranscriptomics.
\newblock \emph{The ISME journal} \textbf{7}, 237 (2013).

\bibitem[{Christiano \emph{et~al.}(2014)Christiano, Nagaraj, Fr{\"o}hlich \&
  Walther}]{christiano2014global}
Christiano, R., Nagaraj, N., Fr{\"o}hlich, F. \& Walther, T.~C.
\newblock Global proteome turnover analyses of the yeasts S. cerevisiae and S.
  pombe.
\newblock \emph{Cell Rep.} \textbf{9}, 1959--1965 (2014).

\bibitem[{Peccoud \& Ycart(1995)}]{peccoud1995markovian}
Peccoud, J. \& Ycart, B.
\newblock Markovian modeling of gene-product synthesis.
\newblock \emph{Theor. Popul. Biol.} \textbf{48}, 222--234 (1995).

\bibitem[{Hornos \emph{et~al.}(2005)}]{hornos2005self}
Hornos, J. \emph{et~al.}
\newblock Self-regulating gene: an exact solution.
\newblock \emph{Phys. Rev. E} \textbf{72}, 051907 (2005).

\bibitem[{Shi \& Qian(2011)}]{shi2011perturbation}
Shi, P.-Z. \& Qian, H.
\newblock A perturbation analysis of rate theory of self-regulating genes and
  signaling networks.
\newblock \emph{J. Chem. Phys.} \textbf{134}, 02B618 (2011).

\bibitem[{Grima \emph{et~al.}(2012)Grima, Schmidt \& Newman}]{grima2012steady}
Grima, R., Schmidt, D. \& Newman, T.
\newblock {Steady-state fluctuations of a genetic feedback loop: An exact
  solution}.
\newblock \emph{J. Chem. Phys.} \textbf{137}, 035104 (2012).

\bibitem[{Potoyan \& Wolynes(2015)}]{potoyan2015dichotomous}
Potoyan, D.~A. \& Wolynes, P.~G.
\newblock Dichotomous noise models of gene switches.
\newblock \emph{J. Chem. Phys.} \textbf{143}, 11B612\_1 (2015).

\bibitem[{Ge \emph{et~al.}(2015)Ge, Qian \& Xie}]{ge2015stochastic}
Ge, H., Qian, H. \& Xie, X.~S.
\newblock Stochastic phenotype transition of a single cell in an intermediate
  region of gene state switching.
\newblock \emph{Phys. Rev. Lett.} \textbf{114}, 078101 (2015).

\bibitem[{Liu \emph{et~al.}(2016)Liu, Yuan, Wang \&
  Zhou}]{liu2016decomposition}
Liu, P., Yuan, Z., Wang, H. \& Zhou, T.
\newblock Decomposition and tunability of expression noise in the presence of
  coupled feedbacks.
\newblock \emph{Chaos: An Interdisciplinary Journal of Nonlinear Science}
  \textbf{26}, 043108 (2016).

\bibitem[{Bintu \emph{et~al.}(2016)}]{bintu2016dynamics}
Bintu, L. \emph{et~al.}
\newblock Dynamics of epigenetic regulation at the single-cell level.
\newblock \emph{Science} \textbf{351}, 720--724 (2016).

\bibitem[{Yin \& Zhang(2013)}]{yin2013continuous}
Yin, G.~G. \& Zhang, Q.
\newblock \emph{{Continuous-time Markov Chains and Applications: A
  Two-time-scale Approach}} (Springer, New York, 2013), 2nd edn.

\bibitem[{Jia(2016)}]{jia2016model}
Jia, C.
\newblock Model simplification and loss of irreversibility.
\newblock \emph{Phys. Rev. E} \textbf{93}, 052149 (2016).

\bibitem[{Chen(2000)}]{chen2000equivalence}
Chen, M.-F.
\newblock Equivalence of exponential ergodicity and L 2-exponential convergence
  for Markov chains.
\newblock \emph{Stochastic processes and their applications} \textbf{87},
  281--297 (2000).

\bibitem[{Chen(2006)}]{chen2006eigenvalues}
Chen, M.-F.
\newblock \emph{{Eigenvalues, Inequalities, and Ergodic Theory}} (Springer,
  2006).

\bibitem[{Meyn \& Tweedie(2012)}]{meyn2012markov}
Meyn, S.~P. \& Tweedie, R.~L.
\newblock \emph{{Markov Chains and Stochastic Stability}} (Springer, 2012).

\bibitem[{Berman \& Plemmons(1979)}]{berman1979nonnegative}
Berman, A. \& Plemmons, R.~J.
\newblock \emph{{Nonnegative Matrices in the Mathematical Sciences}} (Academic
  Press, New York, 1979).

\bibitem[{Chen(1991)}]{chen1991exponentiall}
Chen, M.
\newblock {Exponential $L^2$-convergence and $L^2$-spectral gap for Markov
  processes}.
\newblock \emph{Acta Mathematica Sinica} \textbf{7}, 19--37 (1991).

\bibitem[{Chen(2004)}]{chen2004markov}
Chen, M.
\newblock \emph{From Markov chains to non-equilibrium particle systems} (World
  Scientific, 2004).

\bibitem[{Chen(1996)}]{chen1996estimation}
Chen, M.
\newblock {Estimation of spectral gap for Markov chains}.
\newblock \emph{Acta Mathematica Sinica} \textbf{12}, 337--360 (1996).

\bibitem[{Chen \emph{et~al.}(2017)Chen, Jia \& Jiang}]{chen2017fluctuation}
Chen, Y., Jia, C. \& Jiang, D.-Q.
\newblock Fluctuation-dissipation theorems for inhomogeneous Markov jump
  processes and a biochemical application.
\newblock \emph{J. Math. Phys.} \textbf{58}, 023302 (2017).

\bibitem[{Chen \& Jia(2017)}]{chen2017mathematical}
Chen, X. \& Jia, C.
\newblock Mathematical foundation of nonequilibrium fluctuation-dissipation
  theorems for inhomogeneous diffusion processes with unbounded coefficients.
\newblock \emph{arXiv preprint arXiv:1708.09744}  (2017).

\bibitem[{Yan \& Hsu(2013)}]{yan2013fluctuation}
Yan, C.-C.~S. \& Hsu, C.-P.
\newblock The fluctuation-dissipation theorem for stochastic
  kinetics—Implications on genetic regulations.
\newblock \emph{J. Chem. Phys.} \textbf{139}, 224109 (2013).

\bibitem[{Jia(2017{\natexlab{b}})}]{jia2017nonequilibrium}
Jia, C.
\newblock Nonequilibrium nature of adaptation in bacterial chemotaxis: A
  fluctuation-dissipation theorem approach.
\newblock \emph{Phys. Rev. E} \textbf{95}, 042116 (2017{\natexlab{b}}).

\bibitem[{Assaf \& Meerson(2006)}]{assaf2006spectral}
Assaf, M. \& Meerson, B.
\newblock Spectral theory of metastability and extinction in birth-death
  systems.
\newblock \emph{Phys. Rev. Lett.} \textbf{97}, 200602 (2006).

\bibitem[{Estrada \emph{et~al.}(2016)Estrada, Wong, DePace \&
  Gunawardena}]{estrada2016information}
Estrada, J., Wong, F., DePace, A. \& Gunawardena, J.
\newblock Information integration and energy expenditure in gene regulation.
\newblock \emph{Cell} \textbf{166}, 234--244 (2016).

\bibitem[{Phillips(2015)}]{phillips2015theory}
Phillips, R.
\newblock Theory in Biology: Figure 1 or Figure 7?
\newblock \emph{Trends Cell Biol.} \textbf{25}, 723--729 (2015).

\end{thebibliography}

\end{document}